\documentclass[aps,prc,groupedaddress,showpacs,manuscript]{revtex4}
\usepackage{amssymb}

\usepackage{graphicx}

\begin{document}

\title{Transport model analysis of the transverse momentum and rapidity dependence of pion interferometry at SPS energies}

\author {Qingfeng Li~$\, ^{1}$\footnote{E-mail address: liqf@fias.uni-frankfurt.de}
\email[]{Qi.Li@fias.uni-frankfurt.de}, Marcus Bleicher~$\, ^{2}$,
Xianglei Zhu~$\, ^{1,2,3}$, and Horst St\"{o}cker~$\, ^{1,2}$}
\address{
1) Frankfurt Institute for Advanced Studies (FIAS), Johann Wolfgang Goethe-Universit\"{a}t, Max-von-Laue-Str.\ 1, D-60438 Frankfurt am Main, Germany\\
2) Institut f\"{u}r Theoretische Physik, Johann Wolfgang Goethe-Universit\"{a}t, Max-von-Laue-Str.\ 1, D-60438 Frankfurt am Main, Germany\\
3) Physics Department, Tsinghua University, Beijing 100084, P.R.
China
 }

%\date{\today}

\begin{abstract}
Based on the UrQMD transport model, the transverse momentum and the
rapidity dependence of the Hanbury-Brown-Twiss (HBT) radii $R_L$,
$R_O$, $R_S$ as well as the cross term $R_{OL}$ at SPS energies are
investigated and compared with the experimental NA49 and CERES data.
The rapidity dependence of the $R_L$, $R_O$, $R_S$ is weak while the
$R_{OL}$ is significantly increased at large rapidities and small
transverse momenta. The HBT ``life-time" issue (the phenomenon that
the calculated $\sqrt{R_O^{2}-R_S^{2}}$ value is larger than the
correspondingly extracted experimental data) is also present at SPS
energies.
\end{abstract}

%\textbf{Keywords}:

\pacs{25.75.Gz,25.75.Dw,24.10.Lx} \maketitle

The high temperature phase of Quantum Chromodynamics (QCD), i.e.,
the Quark Gluon Plasma (QGP) has been investigated theoretically
and experimentally for many years with energies in $\sqrt{s}$
ranging from around 2 GeV up to 200 GeV. Since the (phase)
transition from hadrons to quarks might take place only in a small
sub-volume and for a short timespan compared to the overall
process of the heavy ion collisions (HICs), it is an intricate but
essential task to explore the space-time structure of the region
of homogeneity in HICs. Fortunately, a well-established technique
called Hanbury-Brown-Twiss interferometry (HBT)
\cite{HBT54,Goldhaber60,Bauer:1993wq} has been developed, which is
now extensively used in the heavy-ion community to determine the
space-time dimensions of the particle emission source for HICs
with energies from SIS to RHIC
\cite{Lisa:2005dd,Appelshauser:1997rr,Appelshauser:1998vn,Appelshauser:1999in,Appelshauser:2002ps,
Tilsner:2003iv,Appelshauser:2004ys,Bearden:2001sy,Peressounko:2004ha,Antinori:2001yi,
Humanic:2005ye,Adams:2003ra,Bearden:1998aq,Adamova:2002ff,Adamova:2002wi,Chung:2002vk,
Ahle:2002mi,Kolb:2003dz,Tomasik:2002rx,Chajecki:2005iv,Lisa:2005js,Adler:2004rq,Back:2004ug,
Adams:2004yc,Adcox:2002uc,Heinz:2002un,Pratt:2005bt,Lisa:2000hw,
Kniege:2004pt,Soff:2000eh,Soff:2001hc,Zschiesche:2001dx}. Studies
on various species of two-particles other than identical charged
pions have been published or are being pursued (see, e.g. Refs.\
\cite{Lisa:2005dd,Lisa:2005js} and the references therein). After
more than two decades of progress aiming at relativistic HICs,
many basic and important systematic features have been discovered
by femtoscopic measurements at SIS, AGS, SPS, and RHIC
\cite{Lisa:2005dd,Lisa:2005js,Chajecki:2005iv,Adamova:2002ff}: the
dependence on system size, collision centrality, rapidity,
transverse momentum and particle mass (or species).

In our recent work \cite{lqf2006}, we have investigated
theoretically the system size- and the centrality dependence of the
HBT radius parameters $R_L$, $R_O$, and $R_S$ at RHIC energies. It
was found that the calculations are well in line with experimental
data, except that the predicted $R_O$ values in central collisions
(besides the elementary p+p collisions) are slightly larger than the
experimental data. Thus, the $R_o/R_s$ ratio is larger than that
extracted from data. This phenomenon is sometimes dubbed the
``HBT-puzzle" and reported by most of the other models
\cite{Lisa:2005dd,Soff:2001hc,Molnar:2002bz,Hirano:2002ds,Zschiesche:2001dx}.
In Ref. \cite{Lin:2002gc} A Multi-Phase Transport model (AMPT)
provides a good fit to data after including a partonic cascade with
a certain parton-scattering cross section, while in Ref.
\cite{Humanic:2005ye} the Hadronic Rescattering Model (HRM), which
is only based on hadronic rescattering with {\it instantaneous}
collisions, can interpret HBT observables as well. Furthermore, J.
G. Cramer et. al. \cite{Cramer:2004ih} presented one relativisitc
quantum-mechanical treatment of refractive and opacity effects on
two-pion correlations and found a chiral phase transition is
required in the dense medium in order to solve the HBT ``puzzle" at
RHIC, which was further supported by S. Pratt in \cite{Pratt:2005hn}
by considering interactions with mean field in classical
trajectories. Ref. \cite{Pratt:2005bt} argued that the origin of
this ``HBT- puzzle" might be multifaceted. In this work, we focus
our investigation on HICs at lower energies --- the SPS energy
regime
--- partly in order to better understand the origin of the large $R_o/R_s$ in present models.
Furthermore, in this energy region, recently the rapidity
dependence of the HBT-radii of the pion source was investigated
systematically by two experimental collaborations NA49
\cite{Kniege:2004pt,Kniege:2006in} and CERES
\cite{Adamova:2002wi}. It is expected that the rapidity (or
pseudorapidity) dependence of the HBT parameters might provide
essential information on the localization of hot and dense nuclear
matter and hence it is worthwhile to investigate theoretically.

By using the Ultra-relativistic Quantum Molecular Dynamics (UrQMD,
v2.2) transport model (employing hadronic and string degrees of
freedom) (for details, the reader is referred to Refs. \cite{Bass98,
Bleicher99,Bra04,Zhu:2005qa}) and the analyzing program CRAB
(v3.0$\beta$) \cite{Koonin:1977fh, Pratt:1994uf,Pratthome} as tools,
we compare the calculated transverse momentum and rapidity
dependence of HBT-radii with the (preliminary) experimental data at
SPS energies \cite{Adamova:2002wi,Kniege:2004pt,Kniege:2006in}. The
HBT-radii are represented by using the Bertsch-Pratt
three-dimensional convention, which is also called the longitudinal
co-moving system or ``Out-Side-Long" system. In this work, the
three-dimensional correlation function is fit with the standard
Gaussian form:

\begin{equation}
C(q_O,q_S,q_L)=1+\lambda
{\rm exp}(-R_L^2q_L^2-R_O^2q_O^2-R_S^2q_S^2-2R_{OL}^2q_Oq_L), \label{fit1}
\end{equation}
in which $q_i$ and $R_i$ are the components of the pair momentum
difference $\bf{q}=\bf{p}_2-\bf{p}_1$ and the homogeneity length
(the HBT-radii) in the $i$ direction, respectively. The $\lambda$
parameter is the incoherence factor which lies for Bose-Einstein
statistics between 0 (complete coherence) and 1 (complete
incoherence).  Experimentally, the $\lambda$ parameter is affected
by many factors e.g. non-pionic contaminations, (long-lived)
resonances, and details of the Coulomb correction. Theoretically,
the $\lambda$ parameter is mainly suppressed by the presence of
mediate/long-lived resonances since pions from the decay of these
resonances come from a larger source than the directly emitted
pions, which results in an overall suppression of the correlation
function. The $R_{OL}$ represents the cross-term, which is supposed
to vanish at mid-rapidity for symmetric systems. At large
rapidities, $R_{OL}$ deviates from zero
\cite{Chapman:1994yv,Wiedemann:1999qn}, which will be investigated
in this paper as well. In the present UrQMD calculations the
potential interactions are not considered (the cascade mode) due to
the oppressive computing time. Coulomb final state interactions
(FSI) are not taken into account in the program CRAB and the fitting
process.

\begin{figure}
\includegraphics[angle=0,width=0.8\textwidth]{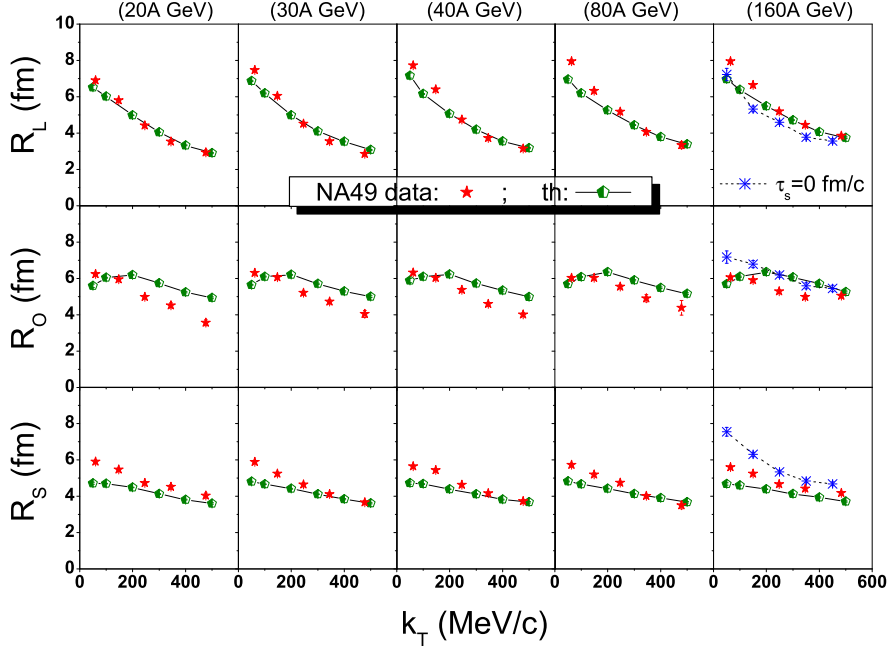}
\caption{(Color Online) Transverse momentum $k_T$ dependence of the
Pratt-radii in central Pb+Pb collisions at $E_b=20$, $30$, $40$,
$80$, and $160$A GeV and mid-rapidity $0<Y_{\pi\pi}<0.5$.
Preliminary NA49 data are taken from \cite{Kniege:2006in}. At
$E_b=160$A GeV the calculation results of the HBT-radii with a
vanishing formation time for strings are also presented. As a
result, the $k_T$-dependence of the HBT radii becomes steeper, and
the values of $R_S$ are increased and approach the calculated $R_O$
values.} \label{fig1}
\end{figure}

\begin{figure}
\includegraphics[angle=0,width=0.8\textwidth]{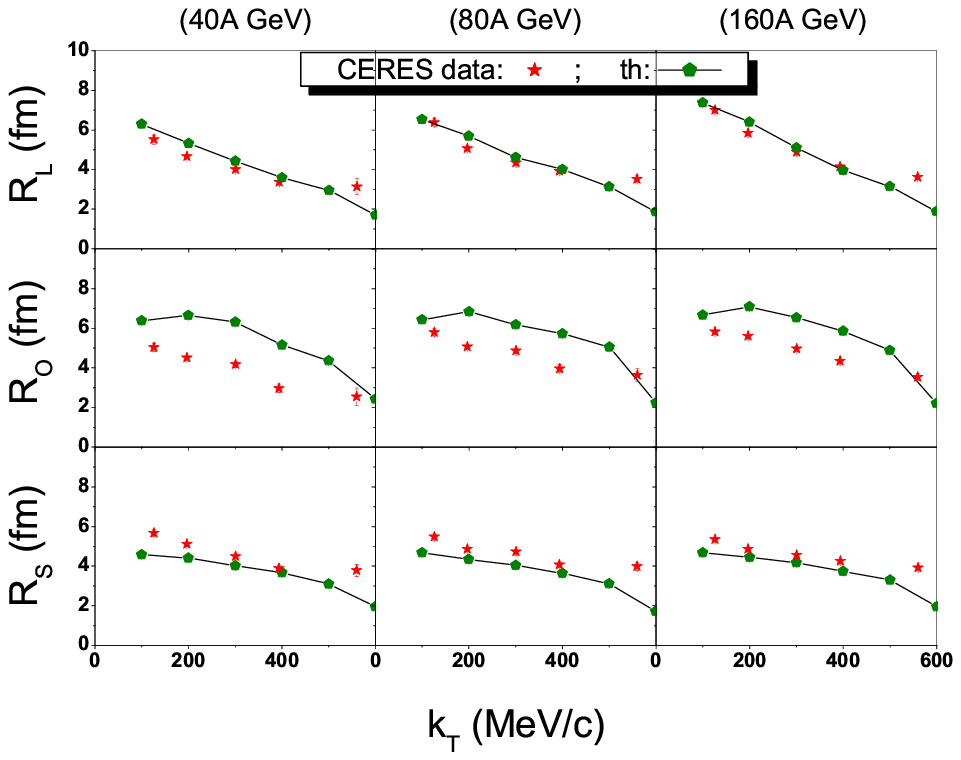}
\caption{(Color Online) Transverse momentum dependence of the
Pratt-radii in central Pb+Au collisions at $E_b=40$
($|Y_{\pi\pi}|<0.25$), $80$ ($-0.5<Y_{\pi\pi}<0$), and $160$A GeV
($-1.0<Y_{\pi\pi}<-0.5$). CERES data are taken from
\cite{Adamova:2002wi}.} \label{fig2}
\end{figure}

In Figs.\ \ref{fig1} and \ref{fig2} we show the transverse momentum
$k_T$ dependence
($\textbf{k}_T=(\textbf{p}_{1T}+\textbf{p}_{2T})/2$) of the HBT-
radii $R_L$ (top plots), $R_O$ (middle plots), and $R_S$ (bottom
plots) in Pb+Pb reactions at beam energies $E_b=20$, $30$, $40$,
$80$, and $160$A GeV, and in Pb+Au reactions at beam energies
$E_b=40$, $80$, and $160$A GeV (plots from left to right) ,
respectively. In Fig.\ \ref{fig1} the results are compared with
preliminary NA49 data \cite{Kniege:2006in} ($<7.2\%$ of the total
cross section $\sigma_T$). The pair-rapidity $0<Y_{\pi\pi}<0.5$
($Y_{\pi\pi}=\frac{1}{2}{\rm log}(\frac{E_1+E_2+p_{\parallel
1}+p_{\parallel 2}}{E_1+E_2-p_{\parallel 1}-p_{\parallel 2}})$ is
the two-pion rapidity with energies $E_1$ and $E_2$ and longitudinal
momenta $p_{\parallel 1}$ and $p_{\parallel 2}$ in the
center-of-mass system) is chosen for all reactions. In Fig.\
\ref{fig2}, the results are compared with CERES data
\cite{Adamova:2002wi} (the $<5\%$ most central collisions). The
$|Y_{\pi\pi}|<0.25$ for 40A GeV, $-0.5<Y_{\pi\pi}<0$ for 80A GeV,
and $-1.0<Y_{\pi\pi}<-0.5$ for 160A GeV, respectively. Furthermore,
the $\pi^{-}-\pi^{-}$ correlations are calculated in Fig.\
\ref{fig1} while the two-charged-pion correlations (including
two-$\pi^-$ and two-$\pi^+$ mesons) are calculated in Fig.\
\ref{fig2} (same as experimental outputs). The statistical errors
are shown as well. The calculated HBT radii at 160 A GeV by adopting
zero formation time for strings ($\tau_s=0$ fm$/$c) will be
explained together with Fig.\ \ref{fig11}. Firstly, it is very
interesting to see that the present calculations can reproduce the
$k_T$-dependence of HBT radii $R_L$ and $R_S$ fairly well shown in
Figs.\ \ref{fig1} and \ref{fig2}. Only at small $k_T$ values, the
calculated $R_L$ and $R_S$ values are seen up to $25\%$ lower than
data. While for the $R_O$ values, the calculations are shown larger
than both NA49 and CERES data especially at relatively large $k_T$.
This deviation appears strongest in most central collisions, which
was also seen at RHIC energies \cite{lqf2006}. By comparing the NA49
data with the CERES data for central Pb+Pb ($\sigma/\sigma_T<7.2\%$)
and central Pb+Au ($\sigma/\sigma_T<5\%$) collisions, one observes
that the CERES $R_O$ data are somewhat smaller than the recent
published NA49 data \cite{Kniege:2006in} especially at large $k_T$
and low beam energy $E_b=40$A GeV although the recent NA49 data at
large $k_T$ have already been driven down visibly when comparing
with those preliminary data in the previous publication
\cite{Kniege:2004pt}. The difference between NA49 and CERES data is
obviously not due to the differently charged pion species since,
similar to the HBT results at RHIC \cite{Adams:2004yc,Adler:2004rq},
we also find that the difference between the HBT-radii of two
positively charged and two negatively charged pions is quite small
at SPS energies. The origin of these differences between data are
still not quite clear.

\begin{figure}
\includegraphics[angle=0,width=0.8\textwidth]{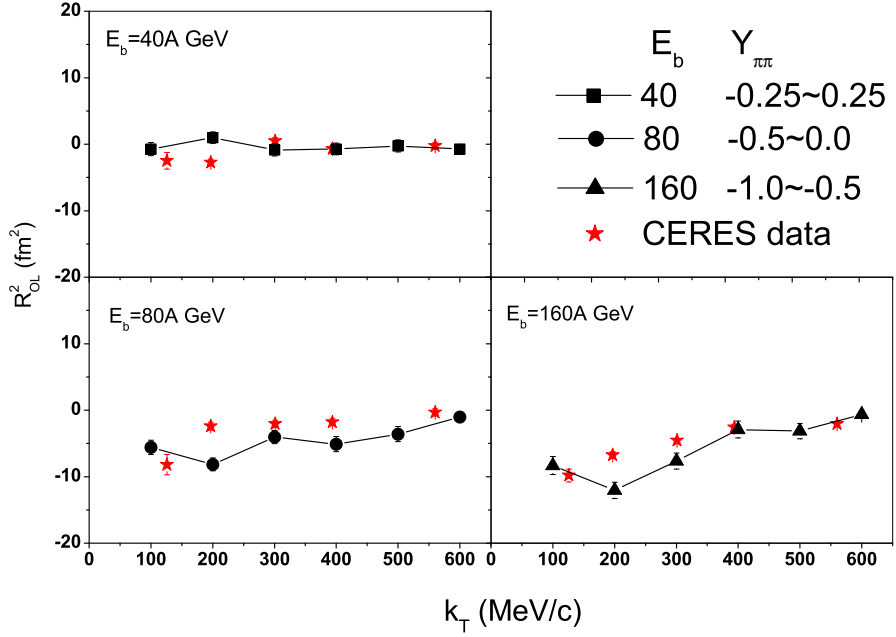}
\caption{(Color Online) $k_T$-dependence of the cross term $R_{OL}^{2}$. Central
Pb+Au collisions at $E_b=40$ ($|Y_{\pi\pi}|<0.25$), $80$
($-0.5<Y_{\pi\pi}<0$), and $160$A GeV ($-1.0<Y_{\pi\pi}<-0.5$) are
chosen for comparing with the CERES data taken from \cite{Adamova:2002wi}.}
\label{fig3}
\end{figure}

In addition to the difference between both experimental data in
centrality, the selected rapidity cuts are also different in both
experiments. The effect of the rapidity-cut on the
$k_T$-dependence of the cross term $R_{OL}^{2}$ is shown in Fig.\
\ref{fig3} in comparison with the CERES data. In line with the
data, an increase of the absolute value of $R_{OL}^{2}$ is seen at
small $k_T$ with the increase of beam energy. This increase is
mainly due to the shift of the rapidity cut away from mid-rapidity
with the increase of the beam energy. This effect will also be
shown in Figs.\ \ref{fig7} and \ref{fig8}. When $k_T$ approaches 0
MeV$/$c, the value of the cross term $R_{OL}^{2}$ also approaches
zero, as one expects \cite{Wiedemann:1999qn}.

\begin{figure}
\includegraphics[angle=0,width=0.8\textwidth]{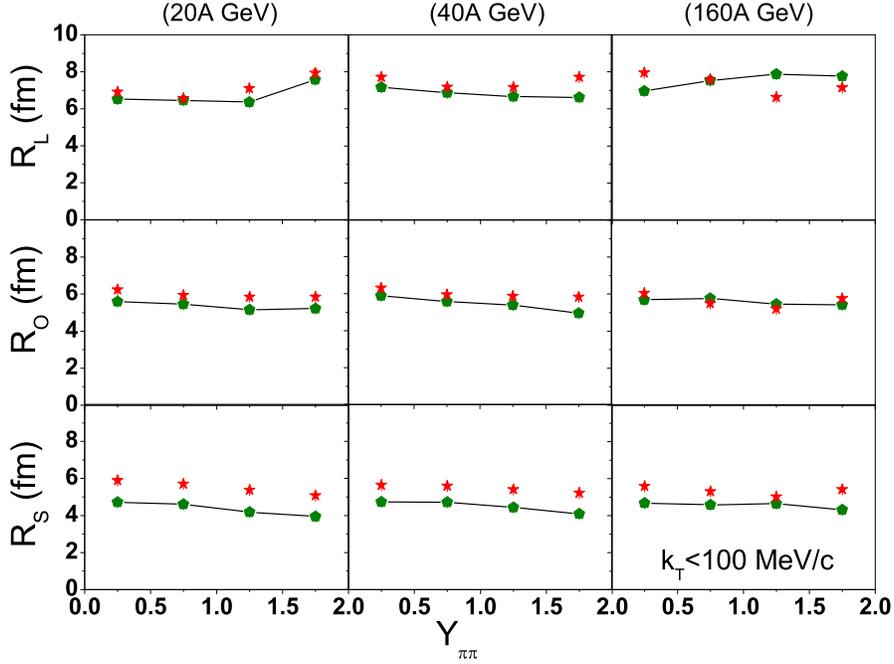}
\caption{(Color Online) Rapidity dependence of the HBT-radii $R_L$
(upper plots), $R_O$ (middle plots), and $R_S$ (bottom plots)
within transverse momentum bin $0<k_T<100 {\rm MeV}/c$  at beam
energies $20$, $40$, and $160$A GeV. The calculations are shown
with lines. Preliminary experimental NA49 data are taken from
\cite{Kniege:2006in} (stars). } \label{fig4}
\end{figure}

\begin{figure}
\includegraphics[angle=0,width=0.8\textwidth]{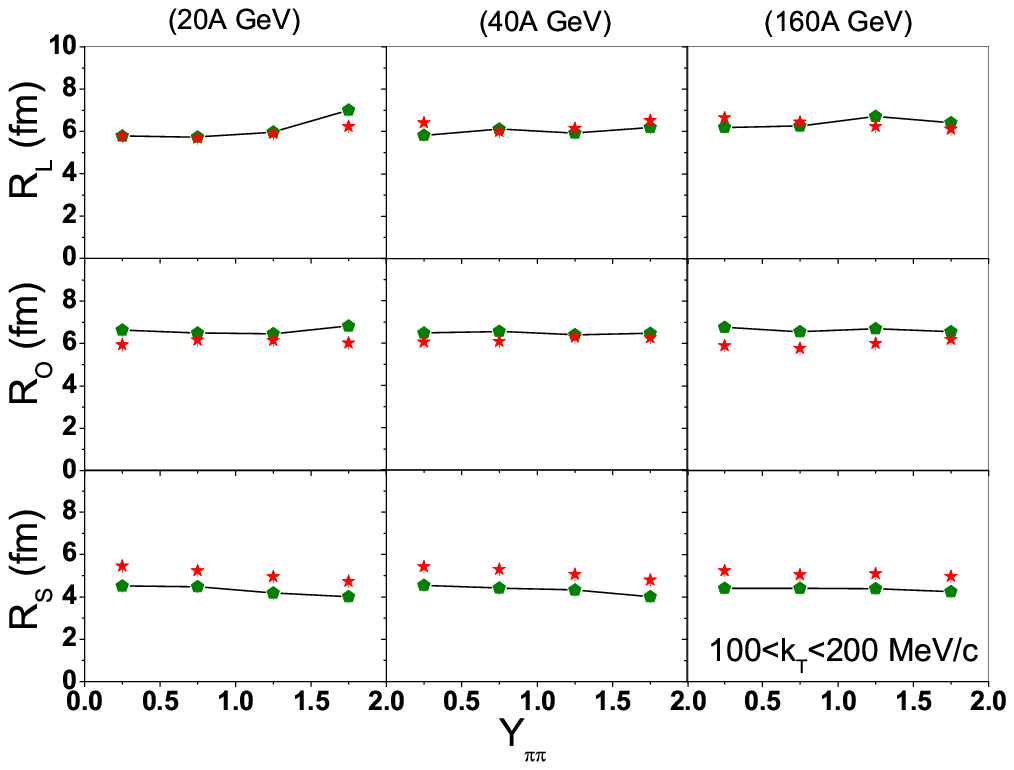}
\caption{(Color Online)  Rapidity dependence of the HBT-radii
$R_L$ (upper plots),  $R_O$ (middle plots), and $R_S$ (bottom
plots) within transverse momentum bin $100<k_T<200 {\rm MeV}/c$
at beam energies $20$, $40$, and $160$A GeV. The calculations are
shown with lines. Preliminary experimental NA49 data are taken
from \cite{Kniege:2006in} (stars). } \label{fig5}
\end{figure}

Preliminary results on the rapidity dependence of $\pi^{-}\pi^{-}$
Bose-Einstein correlations in central Pb+Pb collisions at $20A$,
$30A$, $40A$, $80A$, and $160A$ GeV measured by the NA49
collaboration are available \cite{Kniege:2006in}.  In Figs.\
\ref{fig4} and \ref{fig5} we show the rapidity dependence of the
HBT radii $R_L$ (upper plots), $R_O$ (middle plots), and $R_S$
(bottom plots) within two transverse momentum bins: $0<k_T<100
{\rm MeV}/c$ and $100 {\rm MeV}/c <k_T<200 {\rm MeV}/c$,
respectively. Here  as well as below we only choose three cases
with beam energies $20$, $40$, and $160$A GeV as examples.  For
$R_S$ values shown in Figs.\ \ref{fig4} and \ref{fig5}, the $R_S$
of NA49 data decrease with increasing rapidity slowly which is
also observed in our calculations despite slightly smaller radii.
For $R_O$ results it is seen that the rapidity dependence is
weaker than for $R_S$. The comparison of our calculated results
with NA49 data for the rapidity dependence of the $R_O$ values is
reasonably well, with the deviation up to $20\%$. For $R_L$
results, we find a stronger rapidity dependence in the low
transverse momentum bin $0<k_T<100 {\rm MeV}/c$ while it is
reduced at  large momenta. At transverse momenta $0<k_T<100 {\rm
MeV}/c$ and at low beam energy $E_b=20$A GeV, the $R_L$ value
first drops down and then rises again with the increase of
rapidity, which is observed as well in our calculations. While at
large beam energy $160$A GeV a different calculated rapidity
dependence of $R_L$ values compared with the preliminary
experimental data is found, that is, with the increase of rapidity
the calculated $R_L$ value first rises and then drops while the
trend of experimental data is still preserved similar to the
reactions at lower beam energies although the minimum shifts to a
larger rapidity.

\begin{figure}
\includegraphics[angle=0,width=0.8\textwidth]{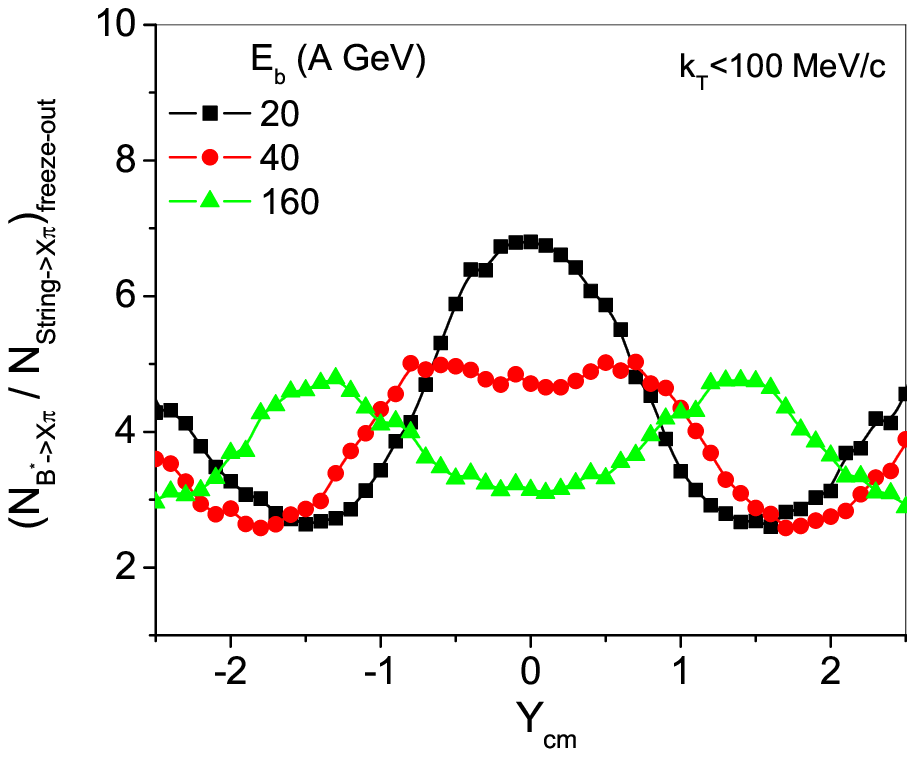}
\caption{(Color Online) Single-particle rapidity ($Y_{\rm cm}$)
distribution of the ratio of the pions at freeze-out from the
decay of resonances and strings.  Three cases at  beam energies
$20$, $40$, and $160$A GeV are chosen. A similar beam-energy
dependence as Fig.\ \ref{fig4} in $R_L$ is seen. (see text for
details) } \label{fig6}
\end{figure}

\begin{figure}
\includegraphics[angle=0,width=0.6\textwidth]{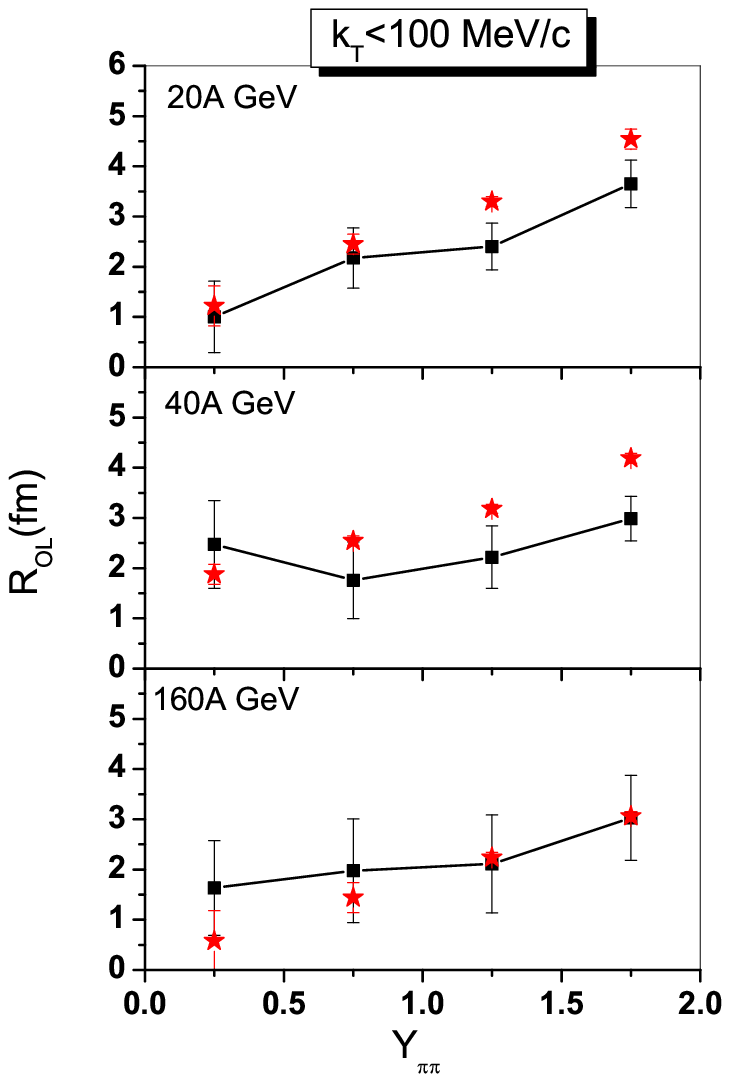}
\caption{(Color Online) Rapidity dependence of $R_{OL}$ calculations (lines) at
transverse momenta $k_T<100 {\rm MeV}/c$ in Pb+Pb reactions at
$20$ (top), $40$ (middle), and $160$A GeV (bottom), compared to
NA49 preliminary data \cite{Kniege:2006in} (stars).} \label{fig7}
\end{figure}

\begin{figure}
\includegraphics[angle=0,width=0.6\textwidth]{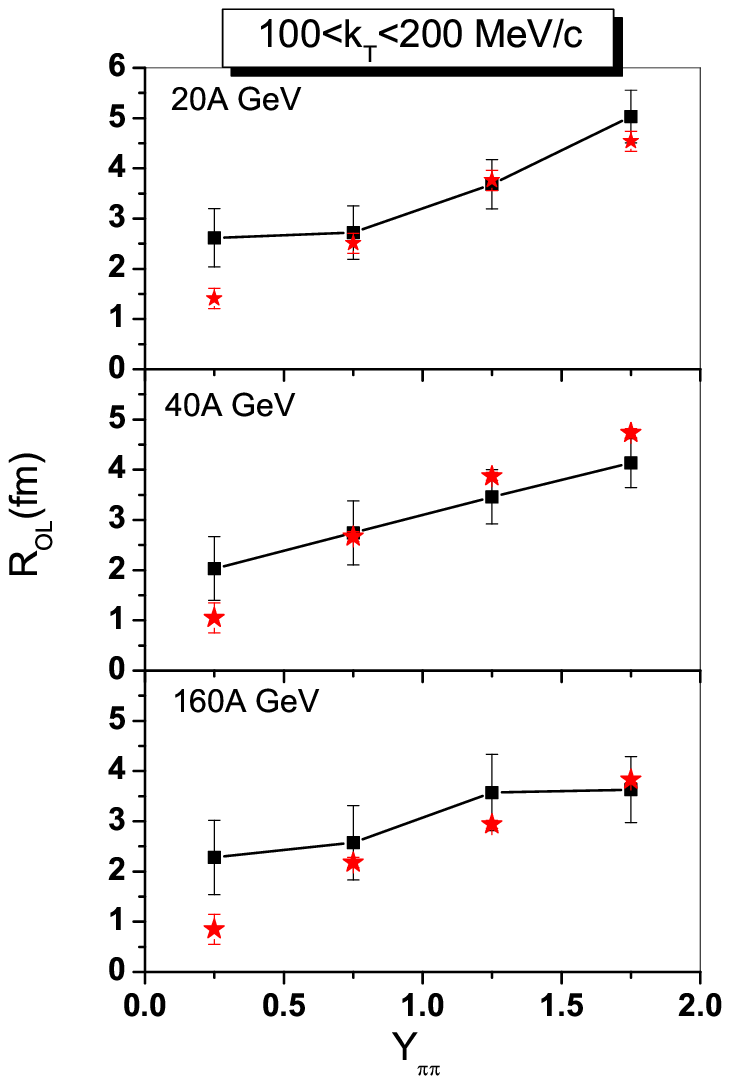}
\caption{(Color Online) Rapidity dependence of $R_{OL}$ calculations (lines) at
transverse momenta $100{\rm MeV}/c <k_T<200 {\rm MeV}/c$ in Pb+Pb reactions at
$20$ (top), $40$ (middle), and $160$A GeV (bottom), compared to
NA49 preliminary data \cite{Kniege:2006in} (stars).} \label{fig8}
\end{figure}

As we know, in UrQMD transport model particles are produced via the
decay of a meson- or baryon-resonance or by color-strings excitation
and fragmentation (the antibaryon-baryon annihilation yields
negligible contributions). Keep in mind that the two ways of
particle production are different: the former is based on the
quasi-particle level (with respect to the resonance masses
$m\lesssim 2.5$ GeV), while the latter happens at large two-particle
center-of-mass energies ($\sqrt{s_{NN}}\gtrsim 2$ GeV). Therefore,
the string process dominates at the early stage of HICs with large
beam energies. While the decay of resonance takes place over the
whole evolution of the system. In order to understand the dissimilar
rapidity-dependence of $R_L$ at large beam energies, we show in
Fig.\ \ref{fig6} the single-pion rapidity ($Y_{cm}=\frac{1}{2}{\rm
log}(\frac{E_1+p_{\parallel 1}}{E_1-p_{\parallel 1}}$)) distribution
of the ratio of the pion numbers at freeze-out from the decay of
resonances and strings. Firstly, we see that the mean ratios have
exceeded unit which means the decay of resonances dominates at
freeze-out and is also of importance to the HBT parameters. In this
figure we also find an interesting beam-energy dependence of the
ratio as a function of rapidity, e.g., at $E_b=20$A GeV, it first
drops from mid-rapidity to projectile-target rapidity region and
then rises further with increasing rapidity. While at the largest
energy point of SPS, $E_b=160$A GeV, the ratio at mid-rapidity is
suppressed due that much more pions produced by string fragmentation
are treated to transversely emit with small longitudinal velocity.
It is known that the case when more pions emitted via the decay of
resonance leads to the postpone of pion freeze-out and hence extend
the volume of source. While larger fraction of pions via the decay
of strings means the freeze-out takes place at earlier stage of
overall process and hence a smaller volume of source. Therefore, the
cancelation of the two mechanisms of pion freeze-out determines the
explicit rapidity dependence of the HBT radii.

Next, we show the rapidity dependence of $R_{OL}$ at transverse
momenta $0<k_T<100 {\rm MeV}/c$ (in Fig.\ \ref{fig7}) and $100{\rm
MeV}/c <k_T<200 {\rm MeV}/c$  (in Fig.\ \ref{fig8}), in Pb+Pb
reactions at $20$ (top), $40$ (middle), and $160$A GeV (bottom). It
is seen clearly that the cross term increases with the increase of
rapidity which is expected since the longitudinally boost-invariance
is broken in the forward and backward rapidity regions
\cite{Wiedemann:1999qn}. We find in our calculations that the
inclusion of this cross term modifies the HBT radii $R_L$, $R_O$,
and $R_S$ only weakly when adopting the rapidity cut
$0<Y_{\pi\pi}<0.5$. However, it is necessary to include this cross
term at large rapidities and small $k_T$ because then it deviates
from zero, as shown in Figs.\ \ref{fig3}, \ref{fig7}, and
\ref{fig8}. One also observes that the present calculations
reproduce the rapidity dependence of $R_{OL}$ values reasonably well
in the rapidity region $0<Y_{\pi\pi}<2$.

\begin{figure}
\includegraphics[angle=0,width=0.6\textwidth]{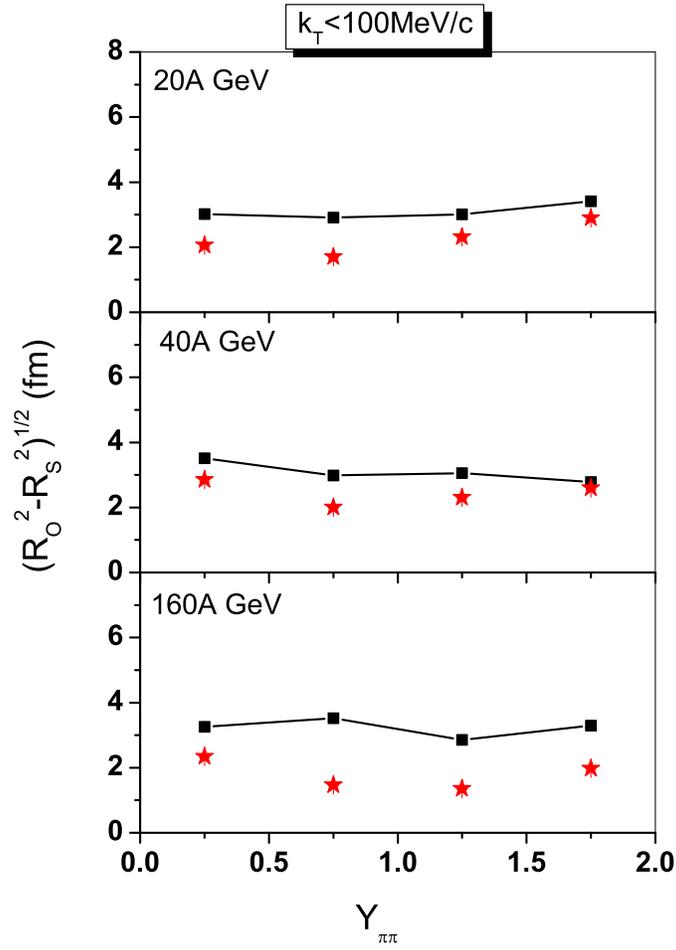}
\caption{(Color Online) Rapidity dependence of the calculated
quantity $\sqrt{R_O^{2}-R_S^{2}}$ (lines) at $k_T<100 {\rm MeV}/c$
with beam energies $20$ (top), $40$ (middle), and $160$A GeV
(bottom). The preliminary experimental data are illustrated with
stars \cite{Kniege:2006in}.} \label{fig9}
\end{figure}

\begin{figure}
\includegraphics[angle=0,width=0.6\textwidth]{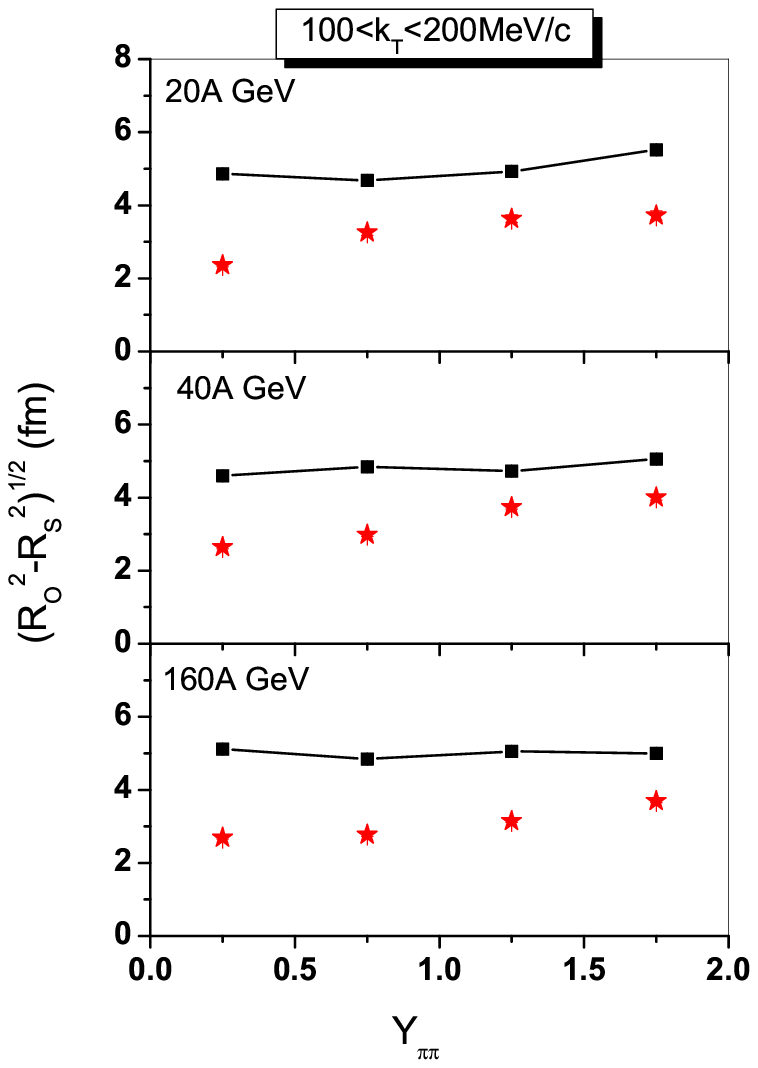}
\caption{(Color Online) Rapidity dependence of the calculated
quantity $\sqrt{R_O^{2}-R_S^{2}}$ (lines) at $100<k_T<200 {\rm
MeV}/c$ with beam energies $20$ (top), $40$ (middle), and $160$A
GeV (bottom). The preliminary experimental data are illustrated
with stars \cite{Kniege:2006in}.} \label{fig10}
\end{figure}

\begin{figure}
\includegraphics[angle=0,width=0.8\textwidth]{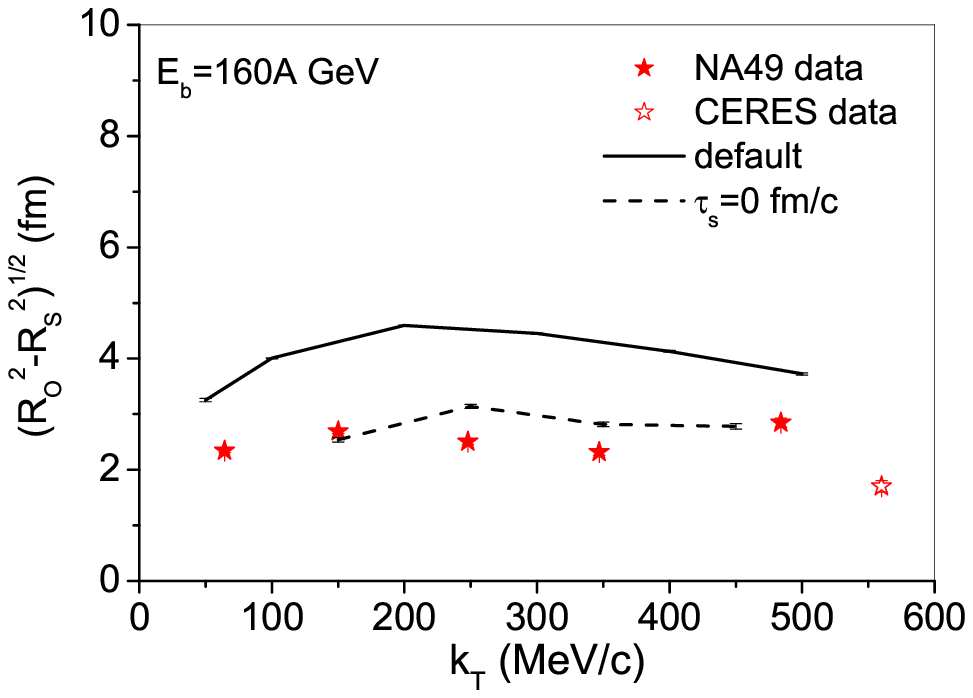}
\caption{(Color Online) $k_T$ dependence of the quantity
$\sqrt{R_O^{2}-R_S^{2}}$ within $0<k_T<100 {\rm MeV}/c$ at beam
energy $160$A GeV. The default calculation (solid line) and the
calculation with a zero formation-time of strings (dashed line) are
compared with NA49 data \cite{Kniege:2006in} (solid stars) at
mid-rapidity as well as CERES data \cite{Adamova:2002wi} (open
stars). The default calculation is somewhat larger than data, which
is similar to those in Figs.\ \ref{fig9} and \ref{fig10}, while the
calculation with reduced formation time matches the measured data on
the ``duration time".} \label{fig11}
\end{figure}

Figs.\ \ref{fig9} and \ref{fig10} show the rapidity dependence of
the quantity $\sqrt{R_O^{2}-R_S^{2}}$ in central $Pb+Pb$ reactions
at $0<k_T<100 {\rm MeV}/c$ and $100<k_T<200 {\rm MeV}/c$,
respectively. In certain scenarios \cite{Heinz:1996qu}, the
so-called ``duration time" $\Delta\tau$ of the pion source is
proportional to this quantity at a fixed transverse momentum and has
been extensively discussed in the HBT community. However, one should
be careful, when interpreting this quantity as ``duration time"
because this identification only holds in the absence of flow and
opacity effects \cite{Heinz:1996qu}. Mainly due to a smaller
calculated $R_S$ in both transverse momentum bins, the calculated
$\sqrt{R_O^{2}-R_S^{2}}$ values are larger than the extracted
experimental data at the rapidity interval considered. It is
interesting to see that the calculation seems to approach the data
at large rapidities $Y_{\pi\pi}=1.5\sim 2.0$. In order to solve the
HBT ``puzzle", the treatment of the interactions during the whole
process of HICs should be important. One should notice that the
largest deviation of the calculated HBT radii from experimental data
comes from the pion correlation at mid-rapidities in the most
central collisions, where the interactions are strongest. Owing to
the omission of treatment on string-string interactions in UrQMD at
early stage, one can argue that the pressure in the early stage
might be too small, resulting in  a delayed source break-up. The
small elliptic flow calculated at RHIC energies is attributed to the
same origin \cite{Zhu:2005qa}. The mean free path of partons (or
formation time of particles) from the hot midrapidity region was
observed closely linked to the strength of the elliptic flow
\cite{Bleicher:2000sx} and should also modify the HBT results. In
Fig.\ \ref{fig1} (in central collisions at $E_b=160$A GeV) we show
the HBT radii with a zero formation time of particles as well. Due
to the reduced mean-free-path of partons and hence much larger
transverse pressure at early stage, the $k_T$ fall-off of the
HBT-radii is seen even stronger than that found in the data which
implies stronger coordinate-momentum correlation in the modified
calculations. The $R_S$ rises more strongly than the other two HBT
radii and approaches the $R_O$ value. In Fig.\ \ref{fig11} the
extracted $\sqrt{R_O^{2}-R_S^{2}}$ values from NA49 as well as from
CERES data for central collisions as a function of $k_T$ at $160$A
GeV are compared with calculations of central $Pb+Pb$ collisions
with the default and the reduced formation times of strings.
Undoubtedly, the reduced formation time of particles leads to
smaller $\sqrt{R_O^{2}-R_S^{2}}$ values. At $k_T<100 {\rm MeV}/c$,
the calculated $R_S$ values are even larger than $R_O$ values. It is
noticed that the CERES $R_S$ data at $k_T\lesssim 500 {\rm MeV}/c$
are also larger than $R_S$ data. It should be noted that, although a
shorter formation time is apt to explain the so-called
``HBT-puzzle", as well the elliptic flow, the absolute values of
HBT-radii are not in line with the data (shown in Fig.\ \ref{fig1}).
A more careful treatment on string-string and/or hadron-string
interactions as well as the decay of resonances are essential to
consistently explain both the single-particle flows and the
two-particle correlations.

To summarize, by using the CRAB program, we analyzed the energy and
the rapidity dependence of the HBT-parameters $R_L$, $R_O$, $R_S$ as
well as the cross term $R_{OL}$ at SPS energies in collisions
calculated within the UrQMD transport approach. As a whole, the
obtained transverse-momentum and rapidity dependences of the
HBT-parameters are in reasonable agreement with the experimental
NA49 and CERES data, except that the calculated $R_O$ at large $k_T$
and $R_S$ at small $k_T$ are somewhat larger than data. The rapidity
dependence of the HBT radii $R_L$, $R_O$, $R_S$ is relatively weak
while that of the cross term $R_{OL}$ is significant. Comparing our
calculations and the experimental data, the so-called ``HBT-puzzle"
reoccurs at SPS energies ---the calculated $\sqrt{R_O^{2}-R_S^{2}}$
values at small $k_T$ are somewhat larger than the experimental
data. This observation might hint to several open questions in the
treatment of the dynamics of HICs, such as the omission of the
string-string interactions, the omission of potential interactions
of hadrons during the evolution and the involved treatment of
resonance production and lifetime in transport models.

\section*{Acknowledgments}
We would like to thank S. Pratt for providing the CRAB program and
acknowledge support by the Frankfurt Center for Scientific Computing
(CSC). We thank H. Appelsh\"{a}user, M. Gyulassy and T.~J.~Humanic
for helpful discussions and valuable suggestions. Q. Li thanks the
Alexander von Humboldt-Stiftung for financial support. This work is
partly supported by GSI, BMBF, DFG, and Volkswagenstiftung.

\end{document}